# Geometric Aspects of the Lame Equation and Plate Theory

Tsai-Jung Chen, Department of Vehicle Engineering, National Pingtung University of Science and Technology, 1, Shuefu Road, Neipu, Pingtung 91201, Taiwan.

Ying-Ji Hong, Department of Mathematics, National Cheng-Kung University, No.1, University Road, Tainan City 70101, Taiwan.
e-mail: yjhong@mail.ncku.edu.tw

## 1. Introduction

  Over the past few years, it is gradually understood that de Rham Cohomology Theory is closely related to Saint-Venant's compatibility condition in the Elasticity Theory [4-6]. In this article, we will discuss the Hodge Theory and de Rham Cohomology Theory hidden in the Lame Equation and in the Plate Theory.

  Although the Kirchhoff-Love Plate Theory and its extension Mindlin-Reissner Plate Theory have been broadly used in the mechanical engineering, civil engineering, and material sciences, it is well-known that these theories are *inconsistent* physical theories [1]. More importantly, it is known that model predictions from these theories are in poor agreement with experimental data [2-3].

  In this article, we will prove a Decomposition Theorem for the solutions of the Lame Equation, and then use this Decomposition Theorem to present a modified Plate Theory, which is *compatible with* the general principles of Physics and the mathematical structure of the Lame Equation. This modified plate theory will be considered as the "physical limit" of the 3-dimensional Lame system of partial differential equations. Theorem 3.1 has been applied to analyze the vibration frequencies of rubber wiper on windshield [17-18].

  The rest of this article is organized as follows.

● In Section 2, we discuss the geometric Theory of Elasticity and the Cauchy dynamic equation for *hyper-elastic* material. Since the Cauchy dynamic equation is *nonlinear* in nature, usually *linearization* of the Cauchy dynamic equation about an *equilibrium state* is adopted in practice. This leads to the vector-valued Lame equation which is central to the applications of the Elasticity Theory to mechanical engineering and material sciences.

● In Section 3, we will prove a general Decomposition Theorem (Theorem 3.1) for the solutions of the Lame equation. Our proof is motivated by the Hodge Theory and the de Rham Cohomology Theory. In this section, general theory of elliptic partial differential equations of second order [14-16] is utilized.

● In Section 4, we will use the Decomposition Theorem (Theorem 3.1) to present an elastic plate theory,which is *compatible with* the general principles of Physics and the mathematical structure of the Lame Equation. We will explain how the Kirchhoff-Love Plate Theory and the Mindlin-Reissner Plate Theory deviate from the "physical limit" of the Elasticity Theory of Continuum Mechanics.

## 2. Three-Dimensional Geometric Theory of Elasticity

In this section, we review briefly the Theory of Elasticity. This Elasticity Theory originated in the general theory of Continuum Mechanics presented by Cauchy [12-13]. For simplicity, we will only dicuss the Elasticity Theory for *hyper-elastic* material, so that a Potential corresponding to the *Stored Energy of Deformation* exists.

According to the Hamilton's Principle, the motion of an elastic body must satisfy the variational formula

$$-\delta \int_{t_i}^{t_f} U_t \cdot dt + \delta \int_{t_i}^{t_f} T_t \cdot dt + \int_{t_i}^{t_f} (\delta W)_t \cdot dt = 0 \qquad (2.1)$$

for any admissible smooth path of virtual deformation. In the Theory of Elasticity, the Potential $U_t$ corresponds to the *Stored Energy* in the elastic body $B_t$ at time $t$. The *virtual work* $(\delta W)_t$ corresponds to the work done by the *Traction* $\tau_t$, along the *boundary surface* $\partial B_t$ of the elastic body $B_t$, on the *virtual deformation* $(\delta x)_t$ at time $t$. We assume, for simplicity, that the *density function* $\rho_t$ of $B_t$ is a constant *independent of t*. It can be shown that the above variational formula is equivalent to

$$\begin{aligned} 0 &= -\delta \int_{t_i}^{t_f} U_t \cdot dt + \delta \int_{t_i}^{t_f} T_t \cdot dt + \int_{t_i}^{t_f} (\delta W)_t \cdot dt \\ &= \int_{t_i}^{t_f} \left( \oiint_{\partial B_t} < -S(n) + \tau_t : (\delta x)_t > \right) \cdot dt + \\ &\quad \int_{t_i}^{t_f} \left( \iiint_{B_t} < -\rho_t \cdot \frac{\partial}{\partial t}\frac{\partial u}{\partial t} + \sum_{k=1}^{3} (\nabla_{e_k} S)(e_k) : (\delta x)_t > \right) \cdot dt + \\ &\quad \int_{t_i}^{t_f} \left( \iiint_{B_t} < F_{ext} : (\delta x)_t > \right) \cdot dt \end{aligned} \qquad (2.2)$$

in which $(\delta x)_t$ is the admissible smooth path of virtual deformation depending on $t$. Here $S$ is the symmetric *Cauchy Stress Tensor*, and $S(n)$ is the vector-valued component of $S$ along the *outer normal* vector field $n$ on the *boundary surface* $\partial B_t$ of the elastic body $B_t$. In (2.2), $u$ is the vector-valued position function of $B_t$. Moreover $e_1, e_2, e_3$ together constitute an orthonormal framing field on $R^3$. Note that

$$\sum_{k=1}^{3} (\nabla_{e_k} S)(e_k) = (\nabla_{e_1} S)(e_1) + (\nabla_{e_2} S)(e_2) + (\nabla_{e_3} S)(e_3) . \qquad (2.3)$$

Here $\nabla_{e_k} S$ is the *Covariant Derivative* of the Cauchy Stress Tensor $S$ along the vector field $e_k$. $F_{ext}$ is the external force (such as gravity) acting on the elastic body $B_t$. In most cases, the external force (such as gravity) is less important, and we may assume, for simplicity, that $F_{ext} = 0$.

Thus the *vector-valued* Cauchy dynamic equation for the elastic body $B_t$ is

$$-\rho \cdot \frac{\partial^2 u}{\partial t^2} + \sum_{k=1}^{3} (\nabla_{e_k} S)(e_k) = 0 \quad \text{on } B_t \tag{2.4}$$

with

$$S(n) = \tau_t \text{ (}traction/stress\text{ given at time } t\text{) on the boundary surface } \partial B_t. \tag{2.5}$$

It should be noted that when some part $Z_o$ of the boundary $\partial B_t$ is *frozen* in the motion of $B_t$, we must have

$$(\delta x)_t = 0 \quad \text{on } Z_o \tag{2.6}$$

and so *no information* about $S(n)$ on $Z_o$ can be inferred from the formula (2.2).

The *vector-valued* Cauchy dynamic equation (2.4) is *nonlinear* in nature. Usually *linearization* of (2.4) about an *equilibrium state* is adopted in practice. Thus we are led to the following *vector-valued* Lame equation for an *isotropic* hyperelastic body

$$\rho \cdot \frac{\partial^2 u}{\partial t^2} = (\lambda + \mu) \cdot \nabla \left( \frac{\partial u_1}{\partial x_1} + \frac{\partial u_2}{\partial x_2} + \frac{\partial u_3}{\partial x_3} \right) + \mu \cdot \left( \frac{\partial^2 u}{\partial x_1^2} + \frac{\partial^2 u}{\partial x_2^2} + \frac{\partial^2 u}{\partial x_3^2} \right). \tag{2.7}$$

with

$$\check{S}(n) = \tau_t \quad \text{(}traction/stress\text{ given at time } t\text{) on the boundary surface } \partial B_t \tag{2.8}$$

in which $\check{S}$ is the *linearization* of $S$ about the same equilibrium state. Here

$$u(t, x) = (u_1(t, x), u_2(t, x), u_3(t, x)) \quad \text{with } x = (x_1, x_2, x_3). \tag{2.9}$$

We may express the *vector-valued* equation (2.7) more explicitly as follows:

$$\rho \cdot \frac{\partial^2 u(t, x)}{\partial t^2} = (\lambda + \mu) \cdot \nabla \left( \frac{\partial u_1(t, x)}{\partial x_1} + \frac{\partial u_2(t, x)}{\partial x_2} + \frac{\partial u_3(t, x)}{\partial x_3} \right) +$$

$$\mu \cdot \left( \frac{\partial^2 u(t,x)}{\partial x_1^2} + \frac{\partial^2 u(t,x)}{\partial x_2^2} + \frac{\partial^2 u(t,x)}{\partial x_3^2} \right)$$

$$= (\lambda + \mu) \cdot \nabla [\operatorname{div} u(t, x)] + \mu \cdot [\Delta u(t, x)]. \tag{2.10}$$

Here $\Delta$ is the *Laplace* operator on the 3-dimensional space $R^3$.

In (2.7), the coefficients $\lambda$ and $\mu$ are called Lame coefficients. Lame coefficients are material constants related to the Young's modulus $E$ and the Poisson's ratio $\sigma$ as follows:

$$\lambda = \frac{\sigma \cdot E}{(1+\sigma) \cdot (1-2\sigma)} \quad \text{and} \quad \mu = \frac{E}{2 \cdot (1+\sigma)}. \tag{2.11}$$

For $j = 1, 2, 3$ and $k = 1, 2, 3$, we define the *strain tensor* as follows:

$$\varepsilon_{jk}(t, \mathbf{x}) = \frac{1}{2}\left(\frac{\partial u_k(t, \mathbf{x})}{\partial x_j} + \frac{\partial u_j(t, \mathbf{x})}{\partial x_k}\right). \tag{2.12}$$

Since the elastic body is *isotropic*, we have the following relation between the *strain tensor* and the symmetric *Cauchy Stress Tensor*

$$<\breve{\mathbf{S}}(\mathbf{e}_j):\mathbf{e}_k> = \lambda \cdot \left(\frac{\partial u_1}{\partial x_1} + \frac{\partial u_2}{\partial x_2} + \frac{\partial u_3}{\partial x_3}\right) \cdot \delta_{jk} + 2\mu \cdot \varepsilon_{jk} \tag{2.13}$$

in which $\delta_{jk} = 1$ for $j = k$, and $\delta_{jk} = 0$ for $j \neq k$. Thus

$$\breve{\mathbf{S}}(\mathbf{n}) = \sum_{k=1}^{3} <\breve{\mathbf{S}}(\mathbf{n}):\mathbf{e}_k> \cdot \mathbf{e}_k = \sum_{k=1}^{3}\left(\sum_{j=1}^{3} <\mathbf{n}:\mathbf{e}_j> \cdot <\breve{\mathbf{S}}(\mathbf{e}_j):\mathbf{e}_k>\right) \cdot \mathbf{e}_k \tag{2.14}$$

can be computed through (2.13) in terms of the partial derivatives of the components of the *displacement* function $\mathbf{u}(t, \mathbf{x})$.

## 3. Decomposition of Solutions of the Lame equation

Assume that $\mathbf{u}(t, \mathbf{x}) = (u_1(t, \mathbf{x}), u_2(t, \mathbf{x}), u_3(t, \mathbf{x}))$ with $\mathbf{x} = (x_1, x_2, x_3)$ is a smooth vector-valued solution for the following *vector-valued* Lame equation

$$\rho \cdot \frac{\partial}{\partial t}\frac{\partial \mathbf{u}}{\partial t} = (\lambda + \mu) \cdot \nabla\left(\frac{\partial u_1}{\partial x_1} + \frac{\partial u_2}{\partial x_2} + \frac{\partial u_3}{\partial x_3}\right) + \mu \cdot \left(\frac{\partial^2 \mathbf{u}}{\partial x_1^2} + \frac{\partial^2 \mathbf{u}}{\partial x_2^2} + \frac{\partial^2 \mathbf{u}}{\partial x_3^2}\right) \tag{3.1}$$

on a hyper-elastic body $\mathbf{B}_0$ (manifold with *boundary*). We define $\theta = \mathrm{div}\,\mathbf{u}$ so that

$$\theta(t, \mathbf{x}) = \left(\frac{\partial u_1(t, \mathbf{x})}{\partial x_1} + \frac{\partial u_2(t, \mathbf{x})}{\partial x_2} + \frac{\partial u_3(t, \mathbf{x})}{\partial x_3}\right). \tag{3.2}$$

It can be inferred from (3.1) readily that $\theta(t, \mathbf{x})$ satisfies

$$\rho \cdot \frac{\partial}{\partial t}\frac{\partial \theta}{\partial t} = \rho \cdot \frac{\partial}{\partial t}\frac{\partial(\mathrm{div}\,\mathbf{u})}{\partial t} = (\lambda + 2\mu) \cdot \left(\frac{\partial^2 \theta}{\partial x_1^2} + \frac{\partial^2 \theta}{\partial x_2^2} + \frac{\partial^2 \theta}{\partial x_3^2}\right)$$

$$= (\lambda + 2\mu) \cdot (\Delta\theta). \tag{3.3}$$

Here $\Delta$ is the *Laplace* operator on the 3-dimensional space $\mathbf{R}^3$.

Now we solve the following elliptic partial differential equation

$$\Delta w = \left(\frac{\partial^2 w}{\partial x_1^2} + \frac{\partial^2 w}{\partial x_2^2} + \frac{\partial^2 w}{\partial x_3^2}\right) = \theta = \mathrm{div}\,\mathbf{u}, \tag{3.4}$$

defined on $\mathbf{B}_0$, depending on time $t$. It follows from the Theory of Partial

Differential Equations [14-16] that a *smooth solution* $w(t, x)$ for (3.4) exists. (However, *uniqueness* of $w(t, x)$ is *not* assured, because we may change the *boundary condition* for $w$.) We define

$$h = \frac{\partial}{\partial t}\frac{\partial w}{\partial t} - \frac{\lambda + 2\mu}{\rho} \cdot \Delta w. \tag{3.5}$$

We infer from (3.3) and (3.4) that

$$\rho \cdot \frac{\partial}{\partial t}\frac{\partial(\Delta w)}{\partial t} = (\lambda + 2\mu) \cdot \Delta(\Delta\theta) \quad \text{or} \quad \Delta\left(\frac{\partial^2 w}{\partial t^2} - \frac{(\lambda + 2\mu)}{\rho} \cdot \Delta w\right) = \Delta h = 0. \tag{3.6}$$

Note that (3.6) means that $h(t, x)$ is a smooth family of *space-harmonic functions* depending on time $t$.

Now we define a smooth family $H(t, x)$ of *space-harmonic functions* depending on time $t$ as follows:

$$H(t, x) = \frac{1}{2} \cdot \int_0^t \left[\int_0^s h(r, x) \cdot dr\right] \cdot ds. \tag{3.7}$$

It can be checked readily that

$$\frac{\partial^2 H(t, x)}{\partial t^2} = h(t, x). \tag{3.8}$$

Thus we infer from (3.5) that

$$\frac{\partial^2 H}{\partial t^2} = h = \frac{\partial^2 w}{\partial t^2} - \frac{\lambda + 2\mu}{\rho} \cdot \Delta w \quad \text{or} \quad 0 = \frac{\partial}{\partial t}\frac{\partial(w - H)}{\partial t} - \frac{\lambda + 2\mu}{\rho} \cdot \Delta(w - H) \tag{3.9}$$

because $H(t, x)$ is a smooth family of *space-harmonic functions*: $\Delta H = 0$.

We define $\hat{w}(t, x) = w(t, x) - H(t, x)$. Then $\hat{w}(t, x)$ solves the following elliptic partial differential equation

$$\Delta \hat{w} = \theta = \text{div } \boldsymbol{u} \tag{3.10}$$

because $\Delta H = 0$. The equation (3.9) means that $\hat{w}(t, x)$ satisfies the wave equation

$$\frac{\partial^2 \hat{w}}{\partial t^2} - \frac{\lambda + 2\mu}{\rho} \cdot (\Delta \hat{w}) = 0 \quad \text{or} \quad \frac{\partial^2 \hat{w}}{\partial t^2} - \frac{\lambda + \mu}{\rho} \cdot [\text{div}(\nabla \hat{w})] - \frac{\mu}{\rho} \cdot (\Delta \hat{w}) = 0. \tag{3.11}$$

We infer from (3.11) that

$$\frac{\partial^2 (\nabla \hat{w})}{\partial t^2} - \frac{\lambda + \mu}{\rho} \cdot \nabla[\text{div}(\nabla \hat{w})] - \frac{\mu}{\rho} \cdot \Delta(\nabla \hat{w}) = 0. \tag{3.12}$$

Note that the equation (3.12) for $\nabla \hat{w}$ is *similar to* the equation (3.1) for $\boldsymbol{u}$.

Now we define

$$\boldsymbol{v}(t, x) = \boldsymbol{u}(t, x) - \nabla \hat{w}(t, x) \quad \text{or} \quad \boldsymbol{v} = (u_1, u_2, u_3) - \left(\frac{\partial \hat{w}}{\partial x_1}, \frac{\partial \hat{w}}{\partial x_2}, \frac{\partial \hat{w}}{\partial x_3}\right). \tag{3.13}$$

By comparing (3.12) and (3.1), we conclude immediately that $\boldsymbol{v} = \boldsymbol{u} - \nabla \hat{w}$ satisfies

$$\frac{\partial^2 v}{\partial t^2} - \frac{\lambda+\mu}{\rho}\cdot\nabla[\operatorname{div} v] - \frac{\mu}{\rho}\cdot\Delta v = 0. \tag{3.14}$$

However, we infer from (3.10) that
$$\operatorname{div} v = \operatorname{div} u - \operatorname{div}\nabla\hat{w} = \operatorname{div} u - \Delta\hat{w} = 0. \tag{3.15}$$

Thus (3.14) is equivalent to
$$\frac{\partial^2 v}{\partial t^2} - \frac{\mu}{\rho}\cdot\Delta v = 0. \tag{3.16}$$

We summarize the results obtained so far in the following

**Theorem 3.1.**  Assume that $u(t,x) = (u_1(t,x), u_2(t,x), u_3(t,x))$ with $x = (x_1, x_2, x_3)$ is a *vector-valued* solution for the following *vector-valued* Lame equation

$$\rho\cdot\frac{\partial^2 u}{\partial t^2} = (\lambda+\mu)\cdot\nabla\left(\frac{\partial u_1}{\partial x_1} + \frac{\partial u_2}{\partial x_2} + \frac{\partial u_3}{\partial x_3}\right) + \mu\cdot\left(\frac{\partial^2 u}{\partial x_1^2} + \frac{\partial^2 u}{\partial x_2^2} + \frac{\partial^2 u}{\partial x_3^2}\right) \tag{3.1}$$

on a hyper-elastic body $B_0$ (manifold with *boundary*). Then the following decomposition

$$u(t,x) = v(t,x) + \nabla\hat{w}(t,x) \tag{3.17}$$

for $u(t,x)$, with $\hat{w}(t,x)$ and $v(t,x)$ respectively satisfying

$$\Delta\hat{w} = \operatorname{div}(\nabla\hat{w}) = \operatorname{div} u \quad\text{and}\quad \operatorname{div} v = 0, \tag{3.18}$$

is possible. Here $\hat{w}(t,x)$ is a smooth function satisfying the wave equation (3.11)

$$\frac{\partial^2 \hat{w}}{\partial t^2} - \frac{\lambda+2\mu}{\rho}\cdot(\Delta\hat{w}) = 0.$$

$v(t,x)$ is a smooth vector-valued function satisfying the wave equation (3.16)

$$\frac{\partial^2 v}{\partial t^2} - \frac{\mu}{\rho}\cdot\Delta v = 0.$$

Note that the decomposition (3.17) is *not* uniquely determined. However, *modulo* the class of smooth *space-harmonic* functions, the decomposition (3.17) can be considered as "unique". ∎

**Remark 3.1.** We explain why the decomposition (3.17) is *not* uniquely determined. In fact, for a space-harmonic function $g(t,x)$ satisfying $\Delta g = 0$, the following expression

$$u(t,x) = [v(t,x) - (t+3)\cdot\nabla g(t,x)] + \nabla[\hat{w}(t,x) + (t+3)\cdot g(t,x)]$$

gives another decomposition for $u(t,x)$. ∎

## 4. Plate Theory

It is usually difficult to solve the Cauchy dynamic equation. Thus numerical methods are frequently adopted for simulation of the Cauchy dynamic equation nowadays. However, for an elastic plate, a simplified version of the Elasticity Theory of Cauchy was presented by the physicist Kirchhoff and mathematician Love.

It was quickly realized that something peculiar exists in the Kirchhoff-Love Plate Theory. In fact, the "free edge boundary condition" of the Kirchhoff-Love Plate Theory does not seem to be reasonable from a physical point of view, and has become a paradox in Plate Theory. Because of this, the "Saint-Venant's Principle" is usually adopted "to *change* the boundary stress/traction condition" for the Kirchhoff-Love dynamic equation

$$\left[ -\frac{\partial^2 G}{\partial t^2} + \frac{h^2}{12} \cdot \frac{\partial^2 (\Delta G)}{\partial t^2} + \frac{-h^2}{12} \cdot \frac{E}{\rho \cdot (1-\sigma^2)} \cdot \Delta(\Delta G) \right] = 0 \tag{4.1}$$

in which $G$ is a function defined on a 2-dimensional region. However, a suitable mathematical version of the "Saint-Venant's Principle" is still not clearly formulated.

Since the model predictions of Kirchhoff-Love dynamic equation are in poor agreement with experimental data, an extension of it, the Mindlin-Reissner Plate Theory [1], was frequently adopted in mechanical engineering, civil engineering, and material sciences [2-3]. However, model predictions of this Mindlin-Reissner Plate Theory are still in poor agreement with experimental data [2-3].

We will present in the following a modified Plate Theory, which is *compatible with* the general principles of Physics and the mathematical structure of the Lame Equation.

Our starting point is the Decomposition Theorem (Theorem 3.1). Assume that

$$\boldsymbol{u}(t,\boldsymbol{x}) = \boldsymbol{v}(t,\boldsymbol{x}) + \nabla \hat{w}(t,\boldsymbol{x}), \tag{4.2}$$

with $\Delta \hat{w} = \mathrm{div}(\nabla \hat{w}) = \mathrm{div}\,\boldsymbol{u}$ and $\mathrm{div}\,\boldsymbol{v} = 0$, is a smooth decomposition of a smooth solution $\boldsymbol{u}(t,\boldsymbol{x})$ for the Cauchy dynamic equation as stated in Theorem 3.1. We will consider the following approximation

$\hat{w}(t,\boldsymbol{x}) \approx \hat{w}_0(t,x_1,x_2) + x_3 \cdot \hat{w}_1(t,x_1,x_2)$ and $\boldsymbol{v}(t,\boldsymbol{x}) \approx \boldsymbol{v}_0(t,x_1,x_2) + x_3 \cdot \boldsymbol{v}_1(t,x_1,x_2)$

satisfying

$$\left( \frac{\partial^2 \hat{w}_0}{\partial x_1^2} + \frac{\partial^2 \hat{w}_0}{\partial x_2^2} \right) = 0 = \frac{\partial^2 \hat{w}_0}{\partial t^2} \tag{4.3}$$

$$\frac{\partial^2 \hat{w}_1}{\partial t^2} - \frac{\lambda + 2\mu}{\rho} \cdot \left( \frac{\partial^2 \hat{w}_1}{\partial x_1^2} + \frac{\partial^2 \hat{w}_1}{\partial x_2^2} \right) = 0 \tag{4.4}$$

$$\boldsymbol{v}_0(t,x_1,x_2) = \left( \frac{\partial V_0}{\partial x_2}, -\frac{\partial V_0}{\partial x_1}, 0 \right) \tag{4.5}$$

$$\frac{\partial^2 V_0}{\partial t^2} - \frac{\mu}{\rho} \cdot \left( \frac{\partial^2 V_0}{\partial x_1^2} + \frac{\partial^2 V_0}{\partial x_2^2} \right) = 0 \tag{4.6}$$

$$\mathbf{v}_1(t, x_1, x_2) = \left( \frac{\partial V_1}{\partial x_2}, -\frac{\partial V_1}{\partial x_1}, 0 \right) \tag{4.7}$$

$$\frac{\partial^2 V_1}{\partial t^2} - \frac{\mu}{\rho} \cdot \left( \frac{\partial^2 V_1}{\partial x_1^2} + \frac{\partial^2 V_1}{\partial x_2^2} \right) = 0. \tag{4.8}$$

To be continued.....